# Novel Theoretical Study of Plasmonic Waves in Graphene-based Cylindrical Waveguides with Gyro-electric Layers

**Mohammad Bagher Heydari** [1,*]

[1,*] School of Electrical Engineering, Iran University of Science and Technology (IUST), Tehran, Iran

[*]Corresponding author: mo_heydari@alumni.iust.ac.ir

**Abstract:** In this paper, we analytically study Surface Plasmon Polaritons (SPPs) in graphene-based cylindrical structures with gyroelectric layers. In the general waveguide, each graphene layer is surrounded by two various gyro-electric materials. New closed-form relations are derived for the field contributions of SPP waves. As a special case, a gyro-electric cylindrical waveguide with double-layer graphene is considered in this article. For the designed structure, the figure of merit (FOM) of 51 is reported for B= 2 T and $\mu_c$=0.7 eV, at the frequency of 33 THz. It is demonstrated that the FOM of the studied waveguide is varied by tuning the chemical potential and the magnetic bias. Harnessing the gyro-electric media together with graphene layers will open a new promising platform for the design and fabrication of novel plasmonic devices in the mid-infrared region.

## 1. Introduction

2-D Materials, crystalline materials with a thickness of single or few-layer atoms, have received an immense interest to extend new devices due to their fascinating features such as high mobility, and high conductivity [1]. Among these 2-D Materials, Graphene, a new-emerging material with a single layer of carbon atoms, has opened a lot of opportunities in optoelectronics, photonics, plasmonics, and bio-sensing [2, 3]. The emergence of graphene has suggested a new platform in "Plasmonics" science due to its notable features [4, 5]. Nowadays, scientists have reported many novel properties for graphene in the THz region, where one of them is certainly the conductivity of the graphene. Based on this tunable parameter, many innovative devices have been designed and fabricated in plasmonics such as sensors [6, 7], couplers [8-10], filters [11-13], resonators [14-16], and circulators [17-20]. Among these devices, graphene-based waveguides play a remarkable role in graphene plasmonics, which are divided into various platforms such as planar [21-33], cylindrical [34-38], and elliptical structures [39-42]. Compared to traditional metal-based plasmonics [43-47], graphene-based devices present high levels of spatial confinements at the THz and mid-infrared regions. The tunability of metal-based plasmonic devices can be acquired by changing the geometry, shape, or size of the dielectric medium. While graphene-based components are tunable via the electrostatic or magnetostatic bias.

In graphene plasmonics, one of the famous ways to efficiently increase the performance of the designed structure is the integration of graphene with other tunable materials such as chiral materials [48-55], and non-linear materials [56-66]. Cylindrical graphene structures, which are investigated in some articles [34, 35, 67-72], are one of the interesting platforms due to their potential applications. In [34], an analytical model is proposed to derive the propagating features and the cut-off wavelength of propagating modes in a graphene-based nano-wire. They showed that field confinement can be obtained by varying the chemical potential of graphene [34]. Yu et al. classified and studied two kinds of complex waves within the graphene-coated silicon nanowire: fast and slow leaky waves [68]. In [71], the authors reported low mode area and long propagation length for propagating plasmons on a hybrid graphene-based cylindrical waveguide. Zhao et al. studied the plasmonic features of graphene-based InGaAs nanowire and reported a high quality-factor for their designed structure [72]. To the best of the author's knowledge, no publication is investigated the plasma multi-layer graphene waveguides. This paper proposes a novel analytical model for these structures, which considers all propagating SPP waves and derives closed-form complicated relations for the field contributions of SPP waves. The proposed, general structure in this paper is constructed of gyro-electric multi-layer layers, where each graphene layer is located between two different gyro-electric materials.

The remainder of the article is organized as follows. In section 2, after introducing the general waveguide, we will propose a mathematical model for it. Then, a special exemplary structure will be studied in section 3. We will show that the propagating properties of the designed waveguide are tunable by the chemical potential and the DC magnetic bias. Finally, the article is concluded in section 4.

## 2. The Proposed General Structure and its Analytical Model

The configuration of the general structure is illustrated in Fig. 1. The general waveguide is formed of various gyro-electric layers, where each gyro-electric layer is surrounded by two different graphene layers. A perpendicular magnetic bias is applied in the z-direction.

The conductivity of the graphene in the *N*-th layer can be modeled by following relation [73]:

$$\sigma_N(\omega,\mu_{c,N},\Gamma_N,T) = \frac{-je^2}{4\pi\hbar} Ln\left[\frac{2|\mu_{c,N}|-(\omega-j2\Gamma_N)\hbar}{2|\mu_{c,N}|+(\omega-j2\Gamma_N)\hbar}\right] + \frac{-je^2 K_B T}{\pi\hbar^2(\omega-j2\Gamma_N)}\left[\frac{\mu_{c,N}}{K_B T}+2Ln\left(1+e^{-\mu_{c,N}/K_B T}\right)\right] \quad (1)$$

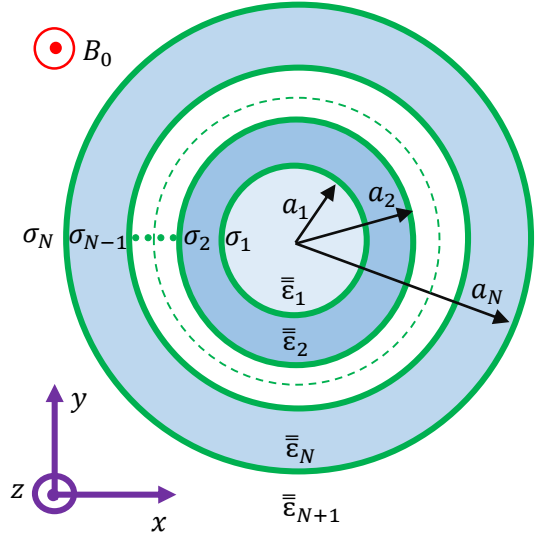

Fig. 1. The cross-section configuration of the graphene-based cylindrical structures with gyro-electric layers.

In (1), $\Gamma_N$ is the scattering rate, $T$ is the temperature, and $\mu_{c,N}$ is the chemical potential for the *N*-th layer [73]. The permittivity tensor of each gyro-electric medium in the *N*-layer can be defined as [74]:

$$\bar{\bar{\varepsilon}}_N = \varepsilon_0 \begin{pmatrix} \varepsilon_N & j\varepsilon_{a,N} & 0 \\ -j\varepsilon_{a,N} & \varepsilon_N & 0 \\ 0 & 0 & \varepsilon_{\|,N} \end{pmatrix} \quad (2)$$

with the following elements [75]:

$$\varepsilon_N = \varepsilon_{\infty,N}\left(1 - \frac{\omega_{p,N}^2(\omega+j\upsilon_N)}{\omega\left[(\omega+j\upsilon_N)^2 - \omega_{c,N}^2\right]}\right) \quad (3)$$

$$\varepsilon_{a,N} = \varepsilon_{\infty,N}\left(\frac{\omega_{p,N}^2 \omega_c}{\omega\left[(\omega+j\upsilon_N)^2 - \omega_c^2\right]}\right) \quad (4)$$

$$\varepsilon_{\parallel,N} = \varepsilon_{\infty,N}\left(1 - \frac{\omega_{p,N}^2}{\omega(\omega + j\upsilon_N)}\right) \tag{5}$$

In (3)-(5), $\nu_N$ is the collision rate and $\varepsilon_{\infty,N}$ is the background permittivity. The plasma and the cyclotron frequency are [75]:

$$\omega_{p,N} = \sqrt{\frac{n_s e^2}{\varepsilon_0 \varepsilon_{\infty,N} m^*}} \tag{6}$$

$$\omega_c = \frac{e B_0}{m^*} \tag{7}$$

By writing Maxwell's equations inside the $N$-th gyro-electric layer (suppose $e^{i\omega t}$) [74]:

$$\nabla \times \mathbf{E} = -j\omega \mu_N \mathbf{H} \tag{8}$$

$$\nabla \times \mathbf{H} = j\omega \bar{\bar{\varepsilon}}_N \cdot \mathbf{E} \tag{9}$$

the z-component of the electromagnetic fields satisfy the following equations [74]:

$$\left(\nabla_\perp^2 + \frac{\varepsilon_{\parallel,N}}{\varepsilon_N}\frac{\partial^2}{\partial z^2} + (k_0^2 \varepsilon_{\parallel,N}\mu_N)\right)E_z + k_0 \mu_N \frac{\varepsilon_{a,N}}{\varepsilon_N}\frac{\partial H_z}{\partial z} = 0 \tag{10}$$

$$\left(\nabla_\perp^2 + \frac{\partial^2}{\partial z^2} + (k_0^2 \varepsilon_{\perp,N}\mu_N)\right)H_z - \frac{k_0 \varepsilon_{\parallel,N}\varepsilon_{a,N}}{\varepsilon_N}\frac{\partial E_z}{\partial z} = 0 \tag{11}$$

where

$$\nabla_\perp^2 = \frac{1}{r}\frac{\partial}{\partial r}r\frac{\partial}{\partial r} + \frac{1}{r^2}\frac{\partial}{\partial^2 \varphi} \tag{12}$$

$$\varepsilon_{\perp,N} = \varepsilon_N - \frac{\varepsilon_{a,N}^2}{\varepsilon_N} \tag{13}$$

The z-components of electromagnetic fields inside the gyro-electric layer can be written as follows (m is an integer):

$$H_{z,N}(r,\varphi,z) = \int_{-\infty}^{+\infty}\sum_{m=-\infty}^{\infty} e^{jk_z z}\exp(-jm\varphi)H_{z,N}^m(\mathrm{r})dk_z \tag{14}$$

$$E_{z,N}(r,\varphi,z) = \int_{-\infty}^{+\infty}\sum_{m=-\infty}^{\infty} e^{jk_z z}\exp(-jm\varphi)E_{z,N}^m(\mathrm{r})dk_z \tag{15}$$

In (14)-(15) the propagation constant is indicated by $k_z$. By substituting (14)-(15) into (10)-(11), we achieve:

$$\left(\nabla_\perp^2 + k_0^2 \varepsilon_{\parallel,N}\mu_N - \frac{\varepsilon_{\parallel,N}}{\varepsilon_N}k_z^2\right)E_{z,N}^m - jk_0 k_z \mu_N \frac{\varepsilon_{a,N}}{\varepsilon_N}H_{z,N}^m = 0 \tag{16}$$

$$\left(\nabla_\perp^2 + k_0^2 \varepsilon_{\perp,N}\mu_N - k_z^2\right)H_{z,N}^m + jk_0 k_z \varepsilon_{\parallel,N}\frac{\varepsilon_{a,N}}{\varepsilon_N}E_{z,N}^m = 0 \tag{17}$$

By combining (16) and (17), the following equation is derived:

$$\left(\nabla_\perp^2 + k_{r,2N-1}^2\right)\left(\nabla_\perp^2 + k_{r,2N}^2\right)H_{z,N}^m = 0 \tag{18}$$

where

$$A_{1,N} = -k_0^2 \mu_N\left(\varepsilon_{\parallel,N} + \varepsilon_{\perp,N}\right) + \left(\frac{\varepsilon_{\parallel,N}}{\varepsilon_N}\right)k_z^2 \tag{19}$$

$$A_{2,N} = \left(k_0^2 \varepsilon_{\|,N} \mu_N - \frac{\varepsilon_{\|,N}}{\varepsilon_N} k_z^2\right)\left(k_0^2 \varepsilon_{\perp,N} \mu_N - k_z^2\right) - k_0^2 k_z^2 \mu_N \varepsilon_{\|,N} \left(\frac{\varepsilon_{a,N}}{\varepsilon_N}\right)^2 \tag{20}$$

Now, the characteristics equation can be expressed as:
$$s^4 + A_{1,N} s^2 + A_{2,N} = 0 \tag{21}$$

with the following roots:

$$k_{r,2N-1} = \sqrt{\frac{-A_{1,N} + \sqrt{A_{1,N}^2 - 4A_{2,N}}}{2}} \tag{22}$$

$$k_{r,2N} = \sqrt{\frac{-A_{1,N} - \sqrt{A_{1,N}^2 - 4A_{2,N}}}{2}} \tag{23}$$

For various regions, the roots can be considered as:

$$k_r = \begin{cases} k_{r,1}, k_{r,2} & i = 1, 2 \ ; \ for\ layer\ N = 1 \\ \ldots & \ldots \\ k_{r,2N-1}, k_{r,2N} & i = 2N-1, 2N \ ; \ for\ layer\ N \\ k_{r,2N+1}, k_{r,2N+2} & i = 2N+1, 2N+2 \ ; \ for\ layer\ N+1 \end{cases} \tag{24}$$

In (24), $N$ is the number of layer and $i$ shows the index of the roots. Now, we write the z-component of the electromagnetic fields in various regions of Fig. 1:

$$H_z^m(r) = \begin{cases} A_{m,1,1} J_m(k_{r,1} r) + A_{m,2,1} J_m(k_{r,2} r) & r < a_1 \\ A_{m,3,2} J_m(k_{r,3} r) + \\ B_{m,3,2} H_m^{(2)}(k_{r,3} r) + \\ A_{m,4,2} J_m(k_{r,4} r) + \\ B_{m,4,2} H_m^{(2)}(k_{r,4} r) & a_1 < r < a_2 \\ \ldots \\ B_{m,2N+1,N+1} H_m^{(2)}(k_{r,2N+1} r) + \\ B_{m,2N+2,N+1} H_m^{(2)}(k_{r,2N+2} r) & r > a_N \end{cases} \tag{25}$$

$$E_z^m(\mathbf{r}) = \begin{cases} A_{m,1,1}T_{1,1}J_m(k_{r,1}r) + A_{m,2,1}T_{2,1}J_m(k_{r,2}r) & r < a_1 \\ T_{3,2}\begin{pmatrix} A_{m,3,2}J_m(k_{r,3}r) + \\ B_{m,3,2}H_m^{(2)}(k_{r,3}r) \end{pmatrix} + \\ T_{4,2}\begin{pmatrix} A_{m,4,2}J_m(k_{r,4}r) + \\ B_{m,4,2}H_m^{(2)}(k_{r,4}r) \end{pmatrix} & a_1 < r < a_2 \\ \ldots \\ T_{2N+1,N+1}B_{m,2N+1,N+1}H_m^{(2)}(k_{r,2N+1}r) + \\ T_{2N+2,N+1}B_{m,2N+2,N+1}H_m^{(2)}(k_{r,2N+2}r) & r > a_N \end{cases} \quad (26)$$

Where

$$T_{i,N} = \frac{1}{-jk_0 k_z \varepsilon_{\|,N}\left(\dfrac{\varepsilon_{a,N}}{\varepsilon_N}\right)}\left(k_{r,i}^2 - k_0^2 \varepsilon_{\perp,N}\mu_N + k_z^2\right) \quad i = 2N-1, 2N \;; N = 1, 2, \ldots \quad (27)$$

The transverse components of electromagnetic fields are obtained as:

$$\begin{pmatrix} E_{r,N} \\ H_{r,N} \end{pmatrix} = \bar{\bar{Q}}_N^{Pos} \frac{\partial}{\partial r}\begin{pmatrix} E_{z,N} \\ H_{z,N} \end{pmatrix} + \frac{m}{r}\bar{\bar{Q}}_N^{Neg}\begin{pmatrix} E_{z,N} \\ H_{z,N} \end{pmatrix} \quad (28)$$

$$j\begin{pmatrix} E_{\varphi,N} \\ H_{\varphi,N} \end{pmatrix} = \bar{\bar{Q}}_N^{Neg} \frac{\partial}{\partial r}\begin{pmatrix} E_{z,N} \\ H_{z,N} \end{pmatrix} + \frac{m}{r}\bar{\bar{Q}}_N^{Pos}\begin{pmatrix} E_{z,N} \\ H_{z,N} \end{pmatrix} \quad (29)$$

In (28)-(29), the following matrices have been utilized:

$$\bar{\bar{Q}}_N^{Pos} = \frac{1}{2}\left[\frac{1}{-k_z^2 + k_0^2 \varepsilon_{+,N}\mu_N}\begin{pmatrix} jk_z & -\omega\mu_0\mu_N \\ \omega\varepsilon_0\varepsilon_{+,N} & jk_z \end{pmatrix} + \frac{1}{-k_z^2 + k_0^2 \varepsilon_{-,N}\mu_N}\begin{pmatrix} jk_z & \omega\mu_0\mu_N \\ -\omega\varepsilon_0\varepsilon_{-,N} & jk_z \end{pmatrix}\right] \quad (30)$$

$$\bar{\bar{Q}}_N^{Neg} = \frac{1}{2}\left[\frac{1}{-k_z^2 + k_0^2 \varepsilon_{+,N}\mu_N}\begin{pmatrix} jk_z & -\omega\mu_0\mu_N \\ \omega\varepsilon_0\varepsilon_{+,N} & jk_z \end{pmatrix} - \frac{1}{-k_z^2 + k_0^2 \varepsilon_{-,N}\mu_N}\begin{pmatrix} jk_z & \omega\mu_0\mu_N \\ -\omega\varepsilon_0\varepsilon_{-,N} & jk_z \end{pmatrix}\right] \quad (31)$$

Moreover,

$$\varepsilon_{\pm,N} = \varepsilon_N \pm \varepsilon_{a,N} \quad (32)$$

By applying the boundary conditions,

$$E_{z,N} = E_{z,N+1}, \; E_{\varphi,N} = E_{\varphi,N+1}$$
(33)

$$H_{z,N+1} - H_{z,N} = -\sigma E_{\varphi,N}, \; H_{\varphi,N+1} - H_{\varphi,N} = \sigma E_{z,N} \quad (34)$$

The final matrix is obtained,

$$\overline{\overline{S}}_{4N,4N} \cdot \begin{pmatrix} A_{m,1,1} \\ A_{m,2,1} \\ ... \\ B_{m,2N+1,N+1} \\ B_{m,2N+2,N+1} \end{pmatrix}_{4N,1} = \begin{pmatrix} 0 \\ 0 \\ ... \\ 0 \\ 0 \end{pmatrix}_{4N,1} \tag{35}$$

In (35), the matrix $\overline{\overline{S}}$ is:

$$\overline{\overline{S}}_{4N,4N4} = \begin{pmatrix} T_{1,1} J_m(k_{r,1} a_1) & ... & ... & 0 & 0 \\ R^J_{1,1,1}(a_1) & ... & ... & 0 & 0 \\ ... & ... & ... & ... & ... \\ 0 & 0 & ... & ... & ... \\ 0 & 0 & ... & ... & -R^H_{2N,N+1,2}(a_N) \end{pmatrix} \tag{36}$$

In (36), the following relations have been utilized,

$$\begin{pmatrix} R^J_{i,N,1}(r) \\ R^J_{i,N,2}(r) \end{pmatrix} = -j \left[ \overline{\overline{Q}}^{Neg}_N k_{r,i} J'_m(k_{r,i} r) + \frac{m}{r} \overline{\overline{Q}}^{Pos}_N J_m(k_{r,i} r) \right] \begin{pmatrix} T_{i,N} \\ 1 \end{pmatrix} \quad i=2N-1, 2N; N=1,2,... \tag{37}$$

$$\begin{pmatrix} R^H_{i,N,1}(r) \\ R^H_{i,N,2}(r) \end{pmatrix} = -j \left[ \overline{\overline{Q}}^{Neg}_N k_{r,i} H'^{(2)}_m(k_{r,i} r) + \frac{m}{r} \overline{\overline{Q}}^{Pos}_N H^{(2)}_m(k_{r,i} r) \right] \begin{pmatrix} T_{i,N} \\ 1 \end{pmatrix} \quad i=2N-1, 2N; N \geq 2 \tag{38}$$

By setting $det(\overline{\overline{S}}) = 0$, the propagation constant and thus other plasmonic features such as the Figure of Merit (FOM) (defined $FOM = Re(k_z)/2\pi Im(k_z)$[76]) can be achieved.

## 3. Results and Discussions

In this section, first, we show the validity of our analytical model, and second, we investigate the performance of a new graphene-based cylindrical waveguide. In all reported results in this section, we assume that the thickness of graphene is $\Delta = 0.5\ nm$, its relaxation time is $\tau = 0.45\ ps$, and the temperature is $T = 300\ K$. To simulate the structures, we have utilized the analysis mode of COMSOL software to obtain the propagation features. The numerical method is the finite element method (FEM). In our simulations, the perfectly matched layers (PML) have been placed around the structures as the boundary conditions. The assumed mesh sizes of the structures are $\Delta x = \Delta y = \Delta z = 0.2$ nm in all directions.

Before embarking on the study of the graphene-based cylindrical waveguide containing a gyro-electric layer, we investigate the accuracy of the model. Consider Fig. 2, where a graphene layer is located on a $SiO_2$ wire, with the permittivity of $\varepsilon_{SiO_2} = 2.09$ and the radius of $R_{SiO_2} = 90\ nm$. FOM is one of the most important factors in studying the performance of plasmonic structures. Fig. 2 illustrates the simulation and analytical results of FOM of a graphene-coated nano-wire as a function of frequency. The results have been depicted for the fundamental mode ($m = 0$) and the first mode ($m = 1$). The analytical and simulation results are prepared by the proposed analytical model and COMSOL software, respectively. In this figure, the SPP wave is a TM mode, because TE plasmonic waves cannot be propagated in this frequency range. FOM increases for the first mode with the increment of frequency but it decreases for the fundamental model. There is an excellent agreement between the analytical and simulation results, which validates the proposed analytical model and shows its high accuracy. Hence, in what follows, we only focus and report the analytical results of the suggested model.

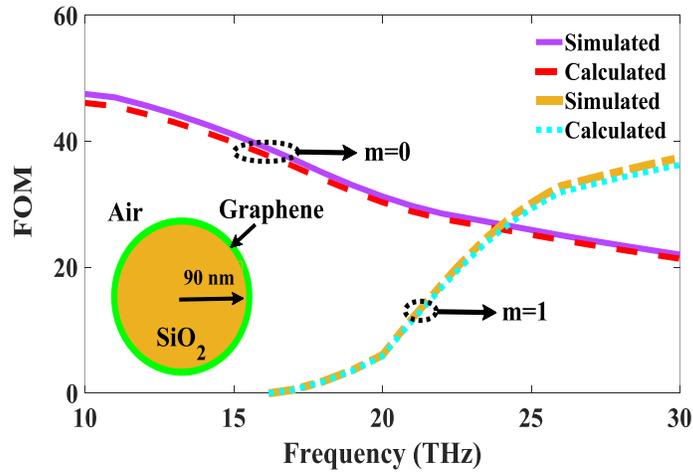

Fig. 2. Comparison of simulation (prepared by COMSOL) and analytical results (prepared by our analytical model) for FOM of a graphene-coated nano-wire as a function of frequency. The results have been depicted for the first two modes. The radius of the $SiO_2$ layer is 90nm with a permittivity of 2.09. The chemical potential is supposed to be 0.45 eV. The relaxation time of the graphene layer is 0.45 ps.

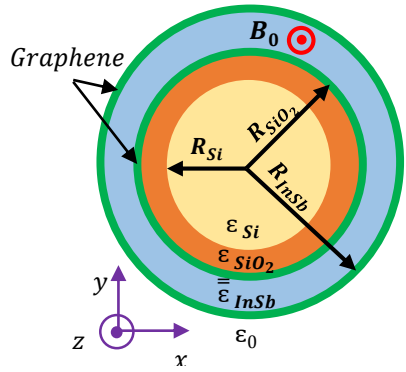

Fig. 3. The cross-section configuration of a cylindrical gyro-electric waveguide with double-layer graphene.

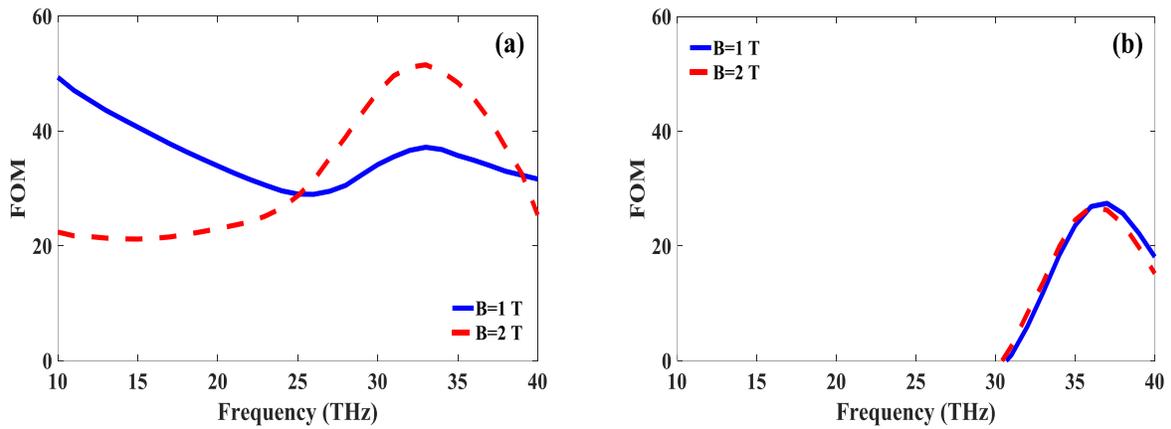

Fig. 4. FOM versus frequency for various magnetic biases ($B_0 = 1, 2$ T) for: **(a)** the fundamental mode (m=0), **(b)** the first mode (m=1)

Fig. 3 shows the cross-section of a new cylindrical gyro-electric waveguide with double-layer graphene, constituting graphene-InSb-graphene-SiO$_2$-Si layers. The permittivity of SiO$_2$ and Si layers are $\varepsilon_{Si} = 11.9$, $\varepsilon_{SiO_2} = 2.09$. The chemical potential is $\mu_c = 0.7\ eV$ unless otherwise stated. The carrier mobility for this structure is supposed to be 6420 cm$^2$/V.s. The plasma layer is the n-type InSb with the following parameters: $\varepsilon_\infty = 15.68, n_s = 1.07 \times 10^{17}/cm^3, \nu = 0.314 \times 10^{13} s^{-1}, m^* = 0.022 m_e$ and $m_e$ is the electron's mass. The geometrical parameters in our results are $t_{InSb} = R_{InSb} - R_{SiO_2} = 5\ nm, R_{Si} = 30\ nm, t_{SiO_2} = R_{SiO_2} - R_{Si} = 3\ nm$, unless otherwise stated.

As mentioned before, FOM is a key parameter for considering the proficiency of the nanostructures. Fig. 4 shows the analytical results of FOM for the cylindrical gyro-electric waveguide with double-layer graphene for various magnetic biases ($B_0 = 1,2\ T$). The results have been depicted for the fundamental mode ($m = 0$) and the first mode ($m = 1$). It should be emphasized that the SPP waves are hybrid TE-TM modes due to the existence of the gyro-electric layer. The fundamental mode ($m = 0$) is cut-off free while the first mode ($m = 1$) has a cut-off frequency (about 31.5 THz). For the fundamental mode ($m = 0$), the FOM for the magnetic bias $B_0 = 1\ T$ is higher than the FOM for the magnetic bias $B_0 = 2\ T$ in the frequency range of $f < 25\ THz$. While the fundamental mode has a higher FOM as the external magnetic bias increases for $f > 25\ THz$. Furthermore, as seen in Fig. 4 (a), there is a maximum point for the FOM diagram for the frequency range of $f > 25\ THz$. For instance, the maximum of FOM for the magnetic bias of $B_0 = 2\ T$ occurs at $f = 33\ THz$ (FOM reaches 51). It is clear from Fig. 4 (b) that the FOM of the first mode ($m = 1$) does not depend strongly on the magnitude of the external magnetic bias.

Fig. 5 demonstrates the FOM as a function of the Si radius ($R_{Si}$) for two modes. In this figure, the magnetic bias is 1 T and the frequency is 35 THz. Other geometrical parameters remained fixed ($t_{SiO_2} = 3\ nm, t_{InSb} = R_{InSb} - R_{SiO_2} = 5\ nm$). As seen in this figure, the first mode has a cut-off radius (18 nm). Therefore, the structure must be designed for $R_{Si} < 18\ nm$ to operate as the single-mode.

In Fig. 6, we have depicted the FOM as a function of the SiO$_2$ thickness ($t_{SiO_2} = R_{SiO_2} - R_{Si}$) for the first two modes. Similar to Fig. 5, the applied bias is 1 T and the operation frequency is supposed to be 35 THz in this figure. Moreover, other geometrical parameters are $R_{Si} = 30\ nm, t_{InSb} = R_{InSb} - R_{SiO_2} = 5\ nm$. One can observe that FOM is very low for $t_{SiO_2} \to 0$. Indeed, for thin thicknesses of SiO$_2$, the propagation length is very low and thus the SPP wave cannot propagate. It is seen from this figure that the fundamental mode has a higher FOM compared to the first mode.

As a final point, the FOM variations have been represented as a function of the chemical potential in Fig. 7. Similar to Fig. 5, 6, the magnetic bias is 1 T and the operation frequency is 35 THz. The geometrical parameters in this figure are $R_{InSb} = 38\ nm, R_{SiO_2} = 33\ nm, R_{Si} = 30\ nm$. As observed in this figure, the FOM for the fundamental mode increases with chemical potential increment. However, for the first mode, there is a maximum point that occurs in the vicinity of 0.7 eV in which the FOM reaches 28 at this point. To fabricate the structure, the SiO$_2$-Si platform is placed on a graphene layer by the micromanipulation technique. Then, an InSb layer is placed on the obtained structure. After that, the structure should be placed on a graphene sheet. Finally, the fabricated structure is eliminated by a tapered fiber from the substrate.

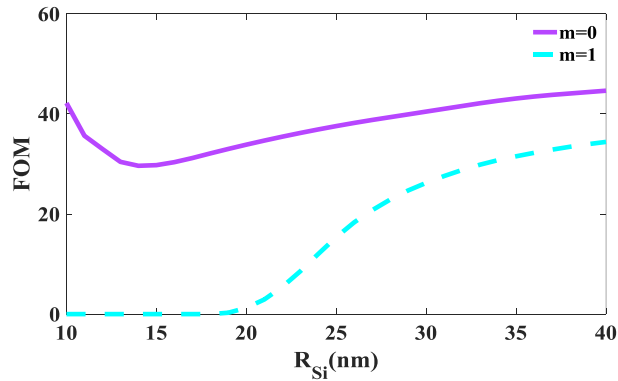

Fig. 5. Dependence of FOM on the Si radius ($R_{Si}$). The frequency is 35 THz.

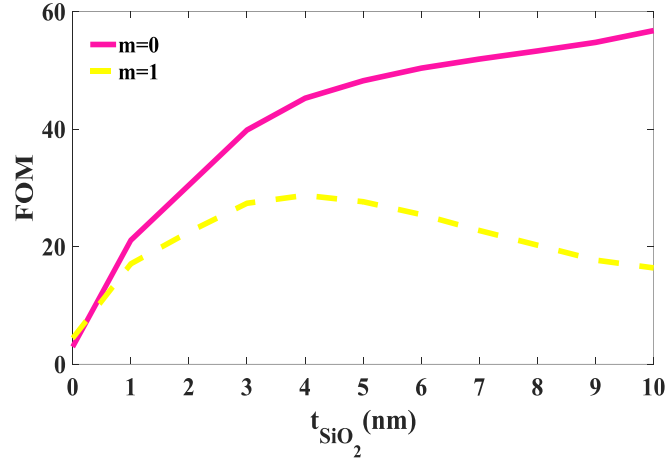

Fig. 6. Dependence of FOM on the SiO$_2$ thickness ($t_{SiO_2} = R_{SiO_2} - R_{Si}$). The frequency is 35 THz.

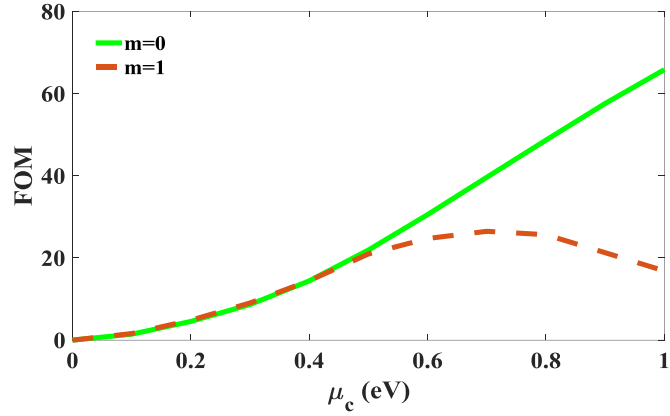

Fig. 7. FOM variations versus the chemical potential ($\mu_c$). The frequency is 35 THz.

## 4. Conclusion

In this study, a new theoretical model was suggested for gyro-electric cylindrical waveguides incorporating graphene layers. To validate the model, the analytical result of FOM was compared to simulation one prepared by COMSOL for a graphene nano-wire. The full agreement between them was seen, which confirmed the high accuracy of our model. As a special case of the general waveguide, a novel gyro-electric-based waveguide with double-layer graphene, constituting graphene-InSb-graphene-SiO$_2$-Si layers, was introduced and investigated. It was shown that the FOM of the designed waveguide could be altered via the chemical potential and the magnetic field.

## References

[1]  P. Avouris, T. F. Heinz, and T. Low, *2D Materials*: Cambridge University Press, 2017.
[2]  V. Singh, D. Joung, L. Zhai, S. Das, S. I. Khondaker, and S. Seal, "Graphene based materials: past, present and future," *Progress in materials science,* vol. 56, pp. 1178-1271, 2011.
[3]  H. Shinohara and A. Tiwari, *Graphene: An Introduction to the Fundamentals and Industrial Applications*: John Wiley & Sons, 2015.
[4]  M. Goerbig, "Electronic properties of graphene in a strong magnetic field," *Reviews of Modern Physics,* vol. 83, p. 1193, 2011.
[5]  M. Sepioni, "Magnetic properties of graphene," 2013.

[6] Y. Zhao, L. Wu, S. Gan, B. Ruan, J. Zhu, X. Dai, *et al.*, "High figure of merit lossy mode resonance sensor with graphene," *Plasmonics,* vol. 14, pp. 929-934, 2019.

[7] X. Yan, T. Wang, X. Han, S. Xiao, Y. Zhu, and Y. Wang, "High Sensitivity Nanoplasmonic Sensor Based on Plasmon-Induced Transparency in a Graphene Nanoribbon Waveguide Coupled with Detuned Graphene Square-Nanoring Resonators," *Plasmonics,* vol. 12, pp. 1449-1455, 2017/10/01 2017.

[8] M. Zhai, H. Peng, X. Wang, X. Wang, Z. Chen, and W. Yin, "The Conformal HIE-FDTD Method for Simulating Tunable Graphene-Based Couplers for THz Applications," *IEEE Transactions on Terahertz Science and Technology,* vol. 5, pp. 368-376, 2015.

[9] M. B. Heydari and M. H. V. Samiei, "Graphene-Based Couplers: A Brief Review," *arXiv preprint arXiv:2010.09462,* 2020.

[10] F. Xu, H. Zhang, and Y. Sun, "Compact graphene directional couplers based on dielectric ridges for planar integration," *Optik-International Journal for Light and Electron Optics,* vol. 131, pp. 588-591, 2017.

[11] D. Correas-Serrano, J. S. Gomez-Diaz, J. Perruisseau-Carrier, and A. Álvarez-Melcón, "Graphene-Based Plasmonic Tunable Low-Pass Filters in the Terahertz Band," *IEEE Transactions on Nanotechnology,* vol. 13, pp. 1145-1153, 2014.

[12] M. B. Heydari and M. H. V. Samiei, "A Short Review on Graphene-Based Filters: Perspectives and Challenges," *arXiv preprint arXiv:2010.07176,* 2020.

[13] H. Zhuang, F. Kong, K. Li, and S. Sheng, "Plasmonic bandpass filter based on graphene nanoribbon," *Applied optics,* vol. 54, pp. 2558-2564, 2015.

[14] T. Christopoulos, O. Tsilipakos, N. Grivas, and E. E. Kriezis, "Coupled-mode-theory framework for nonlinear resonators comprising graphene," *Physical Review E,* vol. 94, p. 062219, 2016.

[15] M. B. Heydari and M. H. V. Samiei, "A Short Review of Plasmonic Graphene-Based Resonators: Recent Advances and Prospects," *arXiv preprint arXiv:2011.14767,* 2020.

[16] X. Zhou, T. Zhang, X. Yin, L. Chen, and X. Li, "Dynamically Tunable Electromagnetically Induced Transparency in Graphene-Based Coupled Micro-ring Resonators," *IEEE Photonics Journal,* vol. 9, pp. 1-9, 2017.

[17] V. Dmitriev, S. L. M. da Silva, and W. Castro, "Ultrawideband graphene three-port circulator for THz region," *Optics express,* vol. 27, pp. 15982-15995, 2019.

[18] M. B. Heydari and M. H. V. Samiei, "Three-port Terahertz Circulator with Multi-layer Triangular Graphene-Based Post," *Optik,* p. 166457, 2021.

[19] V. Dmitriev and W. Castro, "Dynamically controllable terahertz graphene Y-circulator," *IEEE Transactions on Magnetics,* vol. 55, pp. 1-12, 2018.

[20] M. B. Heydari and M. H. V. Samiei, "A Novel Graphene-Based Circulator with Multi-layer Triangular Post for THz Region," *arXiv preprint arXiv:2102.02683,* 2021.

[21] G. W. Hanson, "Quasi-transverse electromagnetic modes supported by a graphene parallel-plate waveguide," *Journal of Applied Physics,* vol. 104, p. 084314, 2008.

[22] M. B. Heydari and M. H. V. Samiei, "Plasmonic graphene waveguides: A literature review," *arXiv preprint arXiv:1809.09937,* 2018.

[23] W. Fuscaldo, P. Burghignoli, P. Baccarelli, and A. Galli, "Complex mode spectra of graphene-based planar structures for THz applications," *Journal of Infrared, Millimeter, and Terahertz Waves,* vol. 36, pp. 720-733, 2015.

[24] M. B. Heydari and M. H. Vadjed Samiei, "An Analytical Study of Magneto-Plasmons in Anisotropic Multi-layer Structures Containing Magnetically Biased Graphene Sheets," *Plasmonics,* vol. 15, pp. 1183-1198, 2020/08/01 2020.

[25] D. Correas-Serrano, J. S. Gomez-Diaz, J. Perruisseau-Carrier, and A. Álvarez-Melcón, "Spatially dispersive graphene single and parallel plate waveguides: Analysis and circuit model," *IEEE Transactions on Microwave Theory and Techniques,* vol. 61, pp. 4333-4344, 2013.

[26] M. B. Heydari and M. H. Vadjed Samiei, "New analytical investigation of anisotropic graphene nano-waveguides with bi-gyrotropic cover and substrate backed by a PEMC layer," *Optical and Quantum Electronics,* vol. 52, p. 108, 2020/02/07 2020.

[27] M. B. Heydari and M. H. V. Samiei, "Analytical Investigation of Magneto-Plasmons in Anisotropic Multi-layer Planar Waveguides Incorporating Magnetically Biased Graphene Sheets," *arXiv preprint arXiv:2103.11452,* 2021.

[28] I. D. Koufogiannis, M. Mattes, and J. R. Mosig, "On the development and evaluation of spatial-domain Green's functions for multilayered structures with conductive sheets," *IEEE Transactions on Microwave Theory and Techniques,* vol. 63, pp. 20-29, 2015.

[29] M. B. Heydari and M. H. V. Samiei, "Magneto-Plasmons in Grounded Graphene-Based Structures with Anisotropic Cover and Substrate," *arXiv preprint arXiv:2103.08557,* 2021.

[30] M. B. Heydari, "Hybrid Graphene-Gyroelectric Structures: A Novel Platform for THz Applications," *arXiv preprint arXiv:2201.06538,* 2022.

[31] M. B. Heydari, "Tunable Plasmonic Modes in Graphene-loaded Plasma Media," 2022.

[32] Y. T. Aladadi and M. A. Alkanhal, "Electromagnetic Characterization of Graphene-Plasma Formations," *IEEE Transactions on Plasma Science,* vol. 48, pp. 852-857, 2020.

[33] M. B. Heydari, "Tunable SPPs supported by hybrid graphene-gyroelectric waveguides: an analytical approach," *Optical and Quantum Electronics,* 2022, https://doi.org/10.1007/s11082-022-03520-2.

[34] Y. Gao, G. Ren, B. Zhu, J. Wang, and S. Jian, "Single-mode graphene-coated nanowire plasmonic waveguide," *Optics letters,* vol. 39, pp. 5909-5912, 2014.

[35] Y. Gao, G. Ren, B. Zhu, H. Liu, Y. Lian, and S. Jian, "Analytical model for plasmon modes in graphene-coated nanowire," *Optics express,* vol. 22, pp. 24322-24331, 2014.

[36] M. B. Heydari and M. H. V. Samiei, "Novel analytical model of anisotropic multi-layer cylindrical waveguides incorporating graphene layers," *Photonics and Nanostructures-Fundamentals and Applications,* vol. 42, p. 100834, 2020.

[37] D. A. Kuzmin, I. V. Bychkov, V. G. Shavrov, and L. N. Kotov, "Transverse-electric plasmonic modes of cylindrical graphene-based waveguide at near-infrared and visible frequencies," *Scientific reports,* vol. 6, p. 26915, 2016.

[38] M. B. Heydari and M. H. V. Samiei, "Anisotropic Multi-layer Cylindrical Structures Containing Graphene Layers: An Analytical Approach," *arXiv preprint arXiv:2103.05594,* 2021.

[39] D. Teng, K. Wang, Z. Li, Y. Zhao, G. Zhao, H. Li*, et al.*, "Graphene-Coated Elliptical Nanowires for Low Loss Subwavelength Terahertz Transmission," *Applied Sciences,* vol. 9, p. 2351, 2019.

[40] M. B. Heydari and M. H. V. Samiei, "A novel analytical study of anisotropic multi-layer elliptical structures containing graphene layers," *IEEE Transactions on Magnetics,* vol. 56, pp. 1-10, 2020.

[41] X. Cheng, W. Xue, Z. Wei, H. Dong, and C. Li, "Mode analysis of a confocal elliptical dielectric nanowire coated with double-layer graphene," *Optics Communications,* vol. 452, pp. 467-475, 2019.

[42] M. B. Heydari and M. H. V. Samiei, "Anisotropic Multi-layer Elliptical Waveguides Incorporating Graphene Layers: A Novel Analytical Model," *arXiv preprint arXiv:2103.01925,* 2021.

[43] S. A. Maier, *Plasmonics: fundamentals and applications*: Springer Science & Business Media, 2007.

[44] F. Kusunoki, T. Yotsuya, J. Takahara, and T. Kobayashi, "Propagation properties of guided waves in index-guided two-dimensional optical waveguides," *Applied Physics Letters,* vol. 86, p. 211101, 2005.

[45] M. B. Heydari, M. Asgari, and N. Jafari, "Novel analytical model for nano-coupler between metal–insulator–metal plasmonic and dielectric slab waveguides," *Optical and Quantum Electronics,* vol. 50, pp. 1-11, 2018.

[46] C. Min and G. Veronis, "Absorption switches in metal-dielectric-metal plasmonic waveguides," *Optics Express,* vol. 17, pp. 10757-10766, 2009.

[47] M. B. Heydari, M. Zolfaghari, M. Asgari, and N. Jafari, "Analysis of Two modified goubau waveguides at THz frequencies: Conical and elliptical structures," *Optik,* vol. 189, pp. 148-158, 2019/07/01/ 2019.

[48] M. Z. Yaqoob, A. Ghaffar, M. Alkanhal, and S. U. Rehman, "Characteristics of light–plasmon coupling on chiral–graphene interface," *JOSA B,* vol. 36, 2019// 2019.

[49] M. B. Heydari and M. H. V. Samiei, "Analytical Study of Chiral Multi-Layer Structures Containing Graphene Sheets for THz Applications," *IEEE Transactions on Nanotechnology,* vol. 19, pp. 653-660, 2020.

[50] M. Yaqoob, A. Ghaffar, M. Alkanhal, S. ur Rehman, and F. Razzaz, "Hybrid Surface Plasmon Polariton Wave Generation and Modulation by Chiral-Graphene-Metal (CGM) Structure," *Scientific reports,* vol. 8, pp. 1-9, 2018.

[51] M. B. Heydari and M. H. V. Samiei, "Chiral Multi-layer Waveguides Incorporating Graphene Sheets: An Analytical Approach," *arXiv preprint arXiv:2102.10135,* 2021.

[52] I. Toqeer, A. Ghaffar, M. Y. Naz, and B. Sultana, "Characteristics of dispersion modes supported by Graphene Chiral Graphene waveguide," *Optik,* vol. 186, 2019// 2019.

[53] M. B. Heydari and M. H. Vadjed Samiei, "Analytical study of hybrid surface plasmon polaritons in a grounded chiral slab waveguide covered with graphene sheet," *Optical and Quantum Electronics,* vol. 52, p. 406, 2020/09/08 2020.

[54] R. Zhao, J. Li, Q. Zhang, X. Liu, and Y. Zhang, "Behavior of SPPs in chiral–graphene–chiral structure," *Optics Letters,* vol. 46, pp. 1975-1978, 2021.

[55] M. B. Heydari and M. H. V. Samiei, "Grounded Graphene-Based Nano-Waveguide with Chiral Cover and Substrate: New Theoretical Investigation," *arXiv preprint arXiv:2102.12465,* 2021.

[56] C. Bhagyaraj, R. Ajith, and M. Vincent, "Propagation characteristics of surface plasmon polariton modes in graphene layer with nonlinear magnetic cladding," *Journal of Optics,* vol. 19, p. 035002, 2017.

[57] M. B. Heydari and M. H. V. Samiei, "Analytical study of TM-polarized surface plasmon polaritons in nonlinear multi-layer graphene-based waveguides," *Plasmonics,* vol. 16, pp. 841-848, 2021.

[58] X. Jiang, J. Gao, and X. Sun, "Control of dispersion properties in a nonlinear dielectric-graphene plasmonic waveguide," *Physica E: Low-dimensional Systems and Nanostructures,* vol. 106, pp. 176-179, 2019.

[59] M. B. Heydari and M. H. V. Samiei, "TM-polarized Surface Plasmon Polaritons in Nonlinear Multi-layer Graphene-Based Waveguides: An Analytical Study," *arXiv preprint arXiv:2101.02536,* 2021.

[60] Y. V. Bludov, D. A. Smirnova, Y. S. Kivshar, N. Peres, and M. I. Vasilevskiy, "Nonlinear TE-polarized surface polaritons on graphene," *Physical Review B,* vol. 89, p. 035406, 2014.

[61] M. B. Heydari, "Analytical Study of TE-Polarized SPPs in Nonlinear Multi-Layer Graphene-Based Structures," *Plasmonics,* vol. 16, pp. 2327-2334, 2021.

[62] Y. Wu, X. Dai, Y. Xiang, and D. Fan, "Nonlinear TE-polarized SPPs on a graphene cladded parallel plate waveguide," *Journal of Applied Physics,* vol. 121, p. 103103, 2017.

[63] M. B. Heydari, "TE-Polarized Surface Plasmon Polaritons (SPPs) in Nonlinear Multi-layer Graphene-Based Waveguides: An Analytical Model," *arXiv preprint arXiv:2107.01684,* 2021.

[64] S. Baher and Z. Lorestaniweiss, "Propagation of surface plasmon polaritons in monolayer graphene surrounded by nonlinear dielectric media," *Journal of Applied Physics,* vol. 124, p. 073103, 2018.

[65] M. B. Heydari, "TE-Polarized SPPs in Nonlinear Multi-layer Graphene-Based Structures," 2021.

[66] W. Walasik, A. Rodriguez, and G. Renversez, "Symmetric plasmonic slot waveguides with a nonlinear dielectric core: Bifurcations, size effects, and higher order modes," *Plasmonics,* vol. 10, pp. 33-38, 2015.

[67] T.-H. Xiao, L. Gan, and Z.-Y. Li, "Graphene surface plasmon polaritons transport on curved substrates," *Photonics Research,* vol. 3, pp. 300-307, 2015.

[68] P. Yu, V. I. Fesenko, and V. R. Tuz, "Dispersion features of complex waves in a graphene-coated semiconductor nanowire," *Nanophotonics,* vol. 7, pp. 925-934, 2018.

[69] Y. Yuan, J. Yao, and W. Xu, "Terahertz photonic states in semiconductor–graphene cylinder structures," *Optics letters,* vol. 37, pp. 960-962, 2012.

[70] D. Correas-Serrano, J. S. Gomez-Diaz, A. Alù, and A. Á. Melcón, "Electrically and Magnetically Biased Graphene-Based Cylindrical Waveguides: Analysis and Applications as Reconfigurable Antennas," *IEEE Transactions on Terahertz Science and Technology,* vol. 5, pp. 951-960, 2015.

[71] J.-P. Liu, X. Zhai, L.-L. Wang, H.-J. Li, F. Xie, Q. Lin*, et al.*, "Analysis of mid-infrared surface plasmon modes in a graphene-based cylindrical hybrid waveguide," *Plasmonics,* vol. 11, pp. 703-711, 2016.

[72] J. Zhao, X. Liu, W. Qiu, Y. Ma, Y. Huang, J.-X. Wang*, et al.*, "Surface-plasmon-polariton whispering-gallery mode analysis of the graphene monolayer coated InGaAs nanowire cavity," *Optics express,* vol. 22, pp. 5754-5761, 2014.

[73] V. Gusynin, S. Sharapov, and J. Carbotte, "Magneto-optical conductivity in graphene," *Journal of Physics: Condensed Matter,* vol. 19, p. 026222, 2006.

[74] A. G. Gurevich and G. A. Melkov, *Magnetization oscillations and waves*: CRC press, 1996.

[75] F. G. Bass and A. A. Bulgakov, *Kinetic and electrodynamic phenomena in classical and quantum semiconductor superlattices*: Nova Publishers, 1997.

[76] P. Berini, "Figures of merit for surface plasmon waveguides," *Optics Express,* vol. 14, pp. 13030-13042, 2006.